\documentstyle[preprint,tighten,eqsecnum,floats,epsfig,aps]{revtex}

\begin{document}

\draft

\title{ Statistical theory of the many-body nuclear system}
\author{ A. De Pace and A. Molinari }
\address{
 Dipartimento di Fisica Teorica dell'Universit\`a di Torino and \\
 Istituto Nazionale di Fisica Nucleare, Sezione di Torino, \\
 via P.Giuria 1, I-10125 Torino, Italy 
}
\date{October 2001}

\maketitle

\begin{abstract}
A recently proposed statistical theory of the mean fields associated with the
ground and excited collective states of a generic many-body system is
extended by increasing the dimensions of the $P$-space. In applying the new
framework to nuclear matter, in addition to the mean field energies we obtain 
their fluctuations as well, together with the ones of the wavefunctions, in
first order of the expansion in the complexity of the $Q$-space states. The
physics described by the latter is assumed to be random. To extract numerical 
predictions out of our scheme we develop a schematic version of the approach,
which, while much simplified, yields results of significance on the size of the
error affecting the mean fields, on the magnitude of the residual effective
interaction, on the ground state spectroscopic factor and on the mixing
occurring between the vectors spanning the $P$-space.
\end{abstract}

\section{ Introduction }
\label{sec:intro}

We have recently addressed the problem of assessing the impact of randomness in
the ground state of atomic nuclei \cite{Fes96,Car98,DeP99}.
In exploring this issue we assume random processes to be related to the strong
repulsion the nucleons experience when they come close to each other inside the
nucleus. This occurrence, while comparatively rare, is of crucial relevance for
the nuclear dynamics and in the perturbative approach one tries to compute the 
related contribution, e.~g., to the binding energy of the system, using 
$G$-matrix techniques.
The latter, however, account solely for the two-body aspect of the short-range
nucleon-nucleon (NN) correlations and do so only within certain approximations.

Furthermore, the large violation of the Hughenoltz-Van Hove theorem occurring
in the Brueckner-Hartree-Fock (BHF) nuclear matter calculations, --- not to 
mention the well-known failure of the BHF theory in accounting for the
empirical data, --- clearly demonstrates that diagrams beyond the ladder ones
are needed to satisfactorily describe the physics of atomic nuclei, in 
particular the short-range one.
Higher order diagrams in the Brueckner-Bethe-Goldstone hole-line expansion have
been computed, but they still appear to fall short of accounting for the
experiments. 

Alternatively, one seeks for an exact solution of the
nuclear many-body problem via the Green's function Monte Carlo method.
This has actually been achieved lately in light nuclei, owing to the remarkable
progress in the computational capabilities of modern computers (see, e.~g.,
Ref.~\cite{Wir00}. 
Yet, even the present days number crunching power is not enough to tackle the
enormously complex problem represented by medium-heavy and heavy nuclei,
precisely as it happens, in a different context, for the non-perturbative QCD.

Moreover, given that these computational obstacles will eventually be
surmounted, one will still have
to face the question whether a non-relativistic potential model description is
sufficient to encompass all the aspects of the nuclear many-body problem.
Indeed, the issue of Lorentz covariance, of particular importance for the
short-range physics in nuclei where large momenta are attained, is lurking in a
corner, not to mention the ambiguities existing in defining and constructing a 
potential model description of the NN interaction at short distances, where it 
is so hard to disentangle particle and nuclear physics from each other and 
where off-shell effects become larger.

It thus appears warranted to explore alternative routes in attacking the
nuclear many-body problem.
Hence, we have recently examined an approach to the theory of the ground
state of atomic nuclei based on the concept of averaging rather than computing 
most of the physics related to the strong NN collisions at short distances.
As a first step we have thus set up an energy averaging procedure suitable for
constructing a mean field to be eventually identified with the shell model.
Here, our procedure parallels the one successfully adopted in the
development of the ``optical model'' of nuclear reactions \cite{Fes92}, to
which it is in fact directly inspired.

Clearly, for the mean field to have a meaning an assessment should be provided
for the error associated with it. Indeed, if the error turns out to be large,
then the concept of mean field is no longer tenable.
Hence, we have derived an expression for the error through an expansion in the
complexity of the states that the NN repulsion at short distances allows to 
reach. At the basis of the derivation lies the
hypothesis alluded to in the beginning of this Introduction, namely that the
matrix elements of the NN repulsion are {\em random}, thus entailing the
vanishing of the average value of the error, but, of course, not of its square.
It is this one that our expansion (finite in finite nuclei and fastly
convergent, as we shall prove) provides.

The above outlined framework basically correspond to a statistical theory of
the ground state (more generally, of the bound states) of the atomic nuclei.
As a first example, we have attempted to implement it in the simplest
among the latter, namely nuclear matter, assuming this system to represent, at
least partially, the physics of heavy nuclei.
Already in performing this task, far from trivial technical problems have been
encountered, notably the one of performing the sum over the
complex nuclear excited states building up the error (square of).
We have been able to surmount this obstacle in Ref.~\cite{DeP99}.

It should be furthermore observed that, at the present stage, even in nuclear 
matter our statistical theory cannot be made parameter--free. 
As a consequence, its predictive power is somewhat limited: Indeed, what we 
actually do amounts to correlate a set of observables of nuclear matter (or 
heavy nuclei), either
measured or computed (via, e.~g., the BHF theory), --- like the binding and the
excited states energies and the level density, --- and ``predict'' on this 
basis the mean field energies of the ground and of the collective nuclear 
states, --- {\em together with the associated error}, --- the residual 
effective interaction, which, while well defined, is presently hardly 
computable, and in addition the ground state spectroscopic factor.

Concerning the parameters entering in our approach, one is needed to 
characterize our energy averaging procedure, whereas the others are required 
whenever the experimental values of the above
referred to observables are lacking or their theoretical evaluation not
available. We shall discuss in particular the significance of the former, which
plays a central role in our approach.

This work is cast in the language introduced to build the 
unified theory of nuclear reactions \cite{Fes92}, at the core of which lies the
partition of the nuclear Hilbert space in the $P$ and $Q$ sectors.
We implement this partition by inserting the random aspects of the physics of
the nucleus in the latter. Hence the $Q$-space should not be viewed as specific
of a given Hamiltonian, but rather it displays universal features and indeed it
is in the $Q$-space that the energy average is performed.
The $P$-space embodies instead the deterministic physics of the nucleus and its
dimensions should be dovetailed for this purpose. This is why the chief scope
of the present research is to expand the dimensions of the $P$ sector of the
nuclear Hilbert space, thus removing a major restriction of our past work,
where the $P$-space  was limited to one dimension only. 

Specifically, the enlargement of the dimensions of the $P$-space allows: i) to
compute the fluctuations of the nuclear wave function and not only of the 
energy, thus assessing the error associated both with the mean field energy 
and its wave function; 
ii) to show the fast convergence of the expansion for the error, essentially
stemming from the rapid increase of the nuclear level density with the 
excitation energy; iii) to provide an improved estimate, with respect to our
past work, of the mean field energy and of the spectroscopic factor of nuclear
matter; iv) to predict a quenching of an order of magnitude of the matrix 
elements of the NN residual effective interaction as compared to those of the
bare force.

In the following we shall dwell on the above items shortly revisiting the
theory in Sect.~\ref{sec:basic}, deepening the meaning of the energy averaging
in Sect.~\ref{sec:eneave} and broadening the $P$-space in
Sect.~\ref{sec:Pspace}. Next, we shall address the question of the
spectroscopic factor and of the wave function fluctuations in 
Sect.~\ref{sec:S}.
Finally, the numerical results obtained on the basis of a schematic model will 
be presented in Sect.~\ref{sec:res}.

\section{ Theoretical framework }
\label{sec:basic}

Our statistical theory has been developed in Refs.~\cite{Fes96,Car98,DeP99}.
For the convenience of the reader and to facilitate the understanding of the
extension of the theory presented in this paper, we shortly recall in
the following the basic equations characterizing our approach.

As it is well-known, the splitting of the Hilbert space induced by the
projection operators $P$ and $Q$ transforms the Schroedinger equation 
\begin{equation}
  \label{eq:Schr}
  H\psi = E\psi
\end{equation}
into the pair of coupled equations
\begin{mathletters}
  \label{eq:Schrpair}
\begin{eqnarray}
  (E-H_{PP})(P\psi) &=& H_{PQ}(Q\psi) \\
  (E-H_{QQ})(Q\psi) &=& H_{QP}(P\psi) ,
\end{eqnarray}
\end{mathletters}
the meaning of the symbols being obvious. From the above the equation obeyed
solely by $(P\psi)$ is derived. It reads
\begin{equation}
  \label{eq:Hcaleq}
  {\cal H}(P\psi) = E (P\psi),
\end{equation}
the $P$-space Hamiltonian being
\begin{equation}
  \label{eq:Hcal}
  {\cal H} = H_{PP}+H_{PQ}
    \frac{1}{\left(\frac{\displaystyle 1}{\displaystyle e_Q}\right)^{-1}+
    W_{QQ}} H_{QP},
\end{equation}
with 
\begin{mathletters}
\begin{eqnarray}
  e_Q    &=& E - H_{QQ} - W_{QQ} \\
  W_{QQ} &=& H_{QP}\frac{1}{E-H_{PP}}H_{PQ}.
\end{eqnarray}
\end{mathletters}
It is of significance that although Eq.~(\ref{eq:Hcaleq}) is not an eigenvalue
equation, since the energy $E$ also appears in the denominator of its right
hand side, yet its solutions only occur for those values of $E$ which are
eigenvalues of (\ref{eq:Schr}) as well; in other words, a one-to-one
correspondence between the values of $E$ allowed by (\ref{eq:Schr}) and
(\ref{eq:Hcaleq}) exists (see next section for a further discussion of this
point).

Now, as already mentioned in the Introduction, in our framework the quantum
deterministic aspect of nuclear dynamics is assumed to be embodied in the
$P$-space, the chaotic one in the $Q$-space. Hence, the strategy of averaging
over the latter follows, although, admittedly, some fuzziness does affect this
partitioning. To set up the averaging procedure we start by the
recognition that the wave functions in the $Q$-space are rapidly varying
functions of the energy $E$, viewed as a parameter classifying their ensemble.

Accordingly, we average over this ensemble following the prescription
\begin{equation}
  \label{eq:aver}
  \langle f(E)\rangle = \int_{-\infty}^{\infty}dE\,\rho(E,\bar{E}_0,\epsilon)
    f(E), 
\end{equation}
$f$ being a generic function to be averaged over the variable $E$ with the
distribution $\rho(E,\bar{E}_0,\epsilon)$.
The latter depends, beyond $E$, also upon the value $\bar{E}_0$ around which
the average, --- taken over a range of $E$ essentially set by $\epsilon$, ---
is performed.
A distribution convenient for our purposes is
\begin{equation}
  \label{eq:rho}
  \rho(E,\bar{E}_0,\epsilon) = \frac{1}{2\pi i}
    \frac{\text{e}^{iE\eta}}{E-(\bar{E}_0-\epsilon)-i\eta} ,
\end{equation}
which is indeed correctly normalized being 
\begin{equation}
  \int_{-\infty}^{\infty}dE\,\rho(E,\bar{E}_0,\epsilon) = 1
\end{equation}
(one should let $\eta\to0^+$ after the integration has been performed).
Note that Eq.~(\ref{eq:rho}) extends in some sense the Lorentz distribution of
the optical model \cite{Fes92} to the situation of a zero width state.
Hence the present formalism is especially suited to deal with the nuclear
ground state, which is of course stable: We shall accordingly focus mainly on
the latter in the following.

We now recall that in \cite{Fes96,Car98,DeP99} the $Q$-space wave functions was
found to read
\begin{equation}
  \label{eq:Qpsi}
  (Q\psi) = \frac{1}{e_Q}H_{QP}\psi_0,
\end{equation}
$\psi_0$ being an auxiliary function that in the end disappears from the
formalism. By averaging Eq.~(\ref{eq:Qpsi}) according to the prescriptions
(\ref{eq:aver}) and (\ref{eq:rho}), one then finds that the energy averaged
wave function of the nuclear ground state in the $P$-space (here denoted by the
angle brackets) obeys the equation
\begin{equation}
  \label{eq:Hbar0eq}
  \bar{\cal H}\langle P\psi\rangle = \bar{E}_0 \langle P\psi\rangle.
\end{equation}
In (\ref{eq:Hbar0eq}) $\bar{E}_0$ is the mean field energy and
\begin{equation}
  \label{eq:Hbar}
  \bar{\cal H} = H_{PP}+H_{PQ}\frac{1}{\left\langle\frac{\displaystyle 1}
    {\displaystyle e_Q}\right\rangle^{-1}+W_{QQ}(E=\bar{E}_0)} H_{QP}
\end{equation}
is the mean field Hamiltonian.
This can be further elaborated since the singularities of the operator $1/e_Q$
lie in the Im$E<0$ half-plane \cite{Bro67}. Accordingly, one gets 
\begin{eqnarray}
  \label{eq:eQaver}
  \langle\frac{1}{e_Q}\rangle &=& \frac{1}{2\pi i}\int_{-\infty}^{\infty}dE
    \frac{\text{e}^{iE\eta}}{E-(\bar{E}_0-\epsilon)-i\eta}
    \frac{1}{E-H_{QQ}-W_{QQ}(E)} \nonumber \\
  &=& \frac{1}{\bar{E}_0-\epsilon-H_{QQ}-W_{QQ}(E=\bar{E}_0-\epsilon)}
    \approx \frac{1}{\bar{E}_0-\epsilon-H_{QQ}-W_{QQ}(E=\bar{E}_0)},
\end{eqnarray}
the last passage holding if the energy dependence of the operator $W_{QQ}$ is 
mild and if the parameter $\epsilon$ is not too large (it should be not too 
small either, otherwise the energy averaging procedure becomes meaningless).

The insertion of (\ref{eq:eQaver}) into (\ref{eq:Hbar}) leads to the following
useful alternative expression for the mean field Hamiltonian
\begin{equation}
  \label{eq:Hbarav}
  \bar{\cal H} = H_{PP} + V_{PQ} V_{QP} \frac{1}{\bar{E}_0-\epsilon-E} ,
\end{equation}
where the energy dependent operators
\begin{mathletters}
  \label{eq:VPQP}
\begin{equation}
  V_{PQ} = H_{PQ}\sqrt{\frac{\bar{E}_0-\epsilon-E}{\bar{E}_0-\epsilon-H_{QQ}}}
\end{equation}
and
\begin{equation}
  V_{QP} = \sqrt{\frac{\bar{E}_0-\epsilon-E}{\bar{E}_0-\epsilon-H_{QQ}}}H_{QP},
\end{equation}
\end{mathletters}
represent the residual effective NN interaction.
The usefulness of the Eqs.~(\ref{eq:Hbarav}) and (\ref{eq:VPQP}) was realized 
in Ref.~\cite{Kaw73}, where it was noticed that with their help the pair of
equations (\ref{eq:Schrpair}) can be recast, as far as $(P\psi)$ is concerned,
into the form
\begin{mathletters}
  \label{eq:Schrav}
\begin{eqnarray}
  \label{eq:Schrava}
  (E-\bar{\cal H})(P\psi) &=& V_{PQ}(Q\psi) \\
  \label{eq:Schravb}
  (E-H_{QQ})(Q\psi)       &=& V_{QP}(P\psi) ,
\end{eqnarray}
\end{mathletters}
which is suitable for expressing the mean field fluctuations (the ``error'').

Indeed, by exploiting the spectral decomposition of the operator 
$(E-\bar{\cal H})^{-1}$  in terms of the eigenfunctions $\phi_n$ of the mean
field Hamiltonian $\bar{\cal H}$, one gets from Eq.~(\ref{eq:Schrava})
\begin{eqnarray}
  \label{eq:spect}
  |P\psi\rangle &=& \sum_n\frac{|\phi_n\rangle}{E-\bar{E}_n}
    \langle\phi_n|V_{PQ}|Q\psi\rangle \nonumber \\
  &=& |\phi_0\rangle\frac{\langle\phi_0|V_{PQ}|Q\psi\rangle}{E-\bar{E}_0} +
    \left(\frac{1}{E-\bar{\cal H}}\right)^\prime V_{PQ}|Q\psi\rangle,
\end{eqnarray}
which, upon left multiplication by $\langle\phi_0|$, yields
\begin{equation}
  \label{eq:phi0ppsi}
  \langle\phi_0|P\psi\rangle = 
    \frac{\langle\phi_0|V_{PQ}|Q\psi\rangle}{E-\bar{E}_0}.
\end{equation}
In the second term on the right hand side of Eq.~(\ref{eq:spect}), the prime
stands for the omission of the $n=0$ term in the spectral decomposition.

Next, the insertion of Eq.~(\ref{eq:spect}) into (\ref{eq:Schravb}) leads to
\begin{equation}
  \label{eq:Qpsipsi}
  |Q\psi\rangle =
   \frac{1}{E-h_{QQ}}V_{QP}|\phi_0\rangle\langle\phi_0|P\psi\rangle,
\end{equation}
where the operator
\begin{equation}
  \label{eq:hQQ}
  h_{QQ} = H_{QQ} + V_{QP}\left(\frac{1}{E-\bar{\cal H}}\right)^\prime V_{PQ}
\end{equation}
has been introduced. Finally, by combining (\ref{eq:phi0ppsi}) and
(\ref{eq:Qpsipsi}), one arrives at the equation
\begin{equation}
  \label{eq:E-E0}
  E-\bar{E}_0 = \langle\phi_0|V_{PQ}\frac{1}{E-h_{QQ}}V_{QP}|\phi_0\rangle,
\end{equation}
which is the basis for computing the mean field energy error.

Although Eq.~(\ref{eq:E-E0}) is valid for any choice of the projectors $P$ and
$Q$, its use is in fact appropriate when the $P$-space is one-dimensional,
as it was indeed the case in our past work, where this choice had been made for
sake of simplicity. For example, 
for a two-dimensional $P$-space, as we shall later see, one should rather 
single out two, rather then one,  terms in the spectral decomposition of the
operator $(E-\bar{\cal H})^{-1}$ on the right hand side of 
Eq.~(\ref{eq:spect}).

We refer the reader to Refs.~\cite{Fes96,Car98,DeP99} for a discussion on how
the average of the square of Eq.~(\ref{eq:E-E0}) is actually computed
(the average of (\ref{eq:E-E0}) of course should vanish) and on how
the complexity expansion is organized. Here, we simply remind that in the
present work, --- as we have done in the past, --- we shall confine
ourselves to the leading term of this fast converging expansion.
The difference between the present results and the previous ones thus
stems entirely from the enlargement of the $P$-space.

\section{ Energy averaging }
\label{sec:eneave}

To understand better the significance of the energy averaging distribution 
(\ref{eq:rho}) let us study how it works in the simple cases of a 
bi-dimensional (A) and of a tri-dimensional (B) Hilbert space.

\subsection{ Bi-dimensional Hilbert space }
\label{subsec:bi}

Let $|\chi_1\rangle$ and $|\chi_2\rangle$ be the two normalized states
spanning the space. Here the only possible choice for the projectors clearly is
\begin{equation}
  P\equiv|\chi_1\rangle\langle\chi_1| \qquad\text{and}\qquad
    Q\equiv|\chi_2\rangle\langle\chi_2|.
\end{equation}
Then, by expanding the operator $1/(E-H_{QQ})$, Eq.~(\ref{eq:Hcaleq}) can be 
recast as follows
\begin{equation}
  \left[E-|\chi_1\rangle a_{11}\langle\chi_1|
    -\chi_1\rangle a_{12}\langle\chi_2|\frac{1}{E}\sum_{n=0}^{\infty}
    \left(\frac{a_{22}}{E}\right)^n(|\chi_2\rangle\langle\chi_2|)^n
    |\chi_2\rangle a_{12}^*\langle\chi_1|\right]|P\psi\rangle=0,
\end{equation}
which, upon multiplying from the left by $\langle\chi_1|$ and exploiting the
idempotency of $|\chi_2\rangle\langle\chi_2|$, simplifies to
\begin{equation}
  \left[E-a_{11}-\frac{|a_{12}|^2}{E-a_{22}}\right]
    \langle\chi_1|P\psi\rangle = 0,
\end{equation}
where the shorthand notations
\begin{equation}
  a_{11}=\langle\chi_1|H|\chi_1\rangle,\quad
  a_{22}=\langle\chi_2|H|\chi_2\rangle \quad\text{and}\quad
  a_{12}=\langle\chi_1|H|\chi_2\rangle
\end{equation}
have been introduced.
This equation is trivially solved yielding the eigenvalues
\begin{equation}
  \label{eq:E+-A}
  E_{\pm} =
   \frac{1}{2}\left[a_{11}+a_{22}\pm\sqrt{(a_{11}-a_{22})^2+4|A_{12}^2}\right],
\end{equation}
which coincide with those of $H$.
It helps notice that the eigenvalues (\ref{eq:E+-A}) are also found as 
intersections of the hyperbola 
\begin{equation}
  \label{eq:hyper}
  E = a_{11} + \frac{|a_{12}|^2}{\omega-a_{22}}
\end{equation}
with the straight line $E=\omega$.

Also the energy averaged Hamiltonian (\ref{eq:Hbarav}) can be expressed in the
basis spanned by $\chi_1$ and $\chi_2$ and one gets the mean field equation
\begin{equation}
  \left[\bar{E}-a_{11}-\frac{|a_{12}|^2}{\bar{E}-a_{22}}\right]
  \langle\chi_1|\biglb\langle P\psi\bigrb\rangle \rangle = 0.
\end{equation}
The latter is again trivially solved yielding
\begin{equation}
  \label{eq:Ebar+-A}
  \bar{E}_{\pm} =\frac{1}{2}\left[a_{11}+a_{22}+\epsilon\pm
    \sqrt{(a_{11}-a_{22}-\epsilon)^2+4b^2}\right],
\end{equation}
which now corresponds to the intersections of the hyperbola (\ref{eq:hyper})
(with $\bar{E}$ replacing $E$) with the new straight line
$\bar{E}=\omega+\epsilon$. 

\begin{figure}[t]
\begin{center}
\mbox{\epsfig{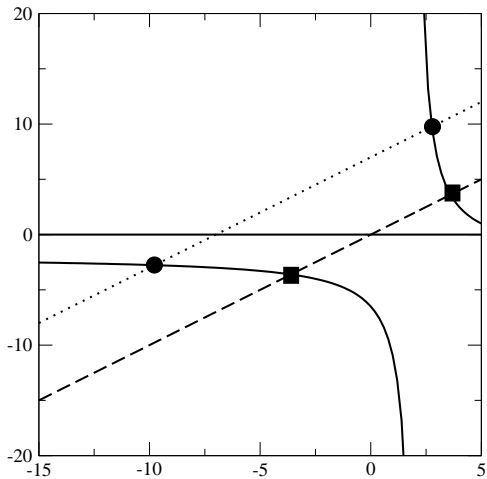}}
\caption{ The eigenvalues of a bi-dimensional Hilbert space. The matrix
elements of the Hamiltonian are taken to be $a_{11}=-2$, $a_{22}=2$ and 
$a_{12}=3$, in arbitrary units.
The exact eigenvalues (squares) and the ones of the energy averaged Hamiltonian
$\bar{\cal H}$ (circles) are shown. They correspond to the intersections with
the straight line $E=\omega+\epsilon$: The former with $\epsilon=0$, the latter
with $\epsilon=7$. The stability of the lowest eigenvalue and the upward shift
of the highest one are clearly apparent. }
\label{fig:hyperbola1}
\end{center}
\end{figure}

From Fig.~\ref{fig:hyperbola1}, where the solutions $E_{\pm}$ and
$\bar{E}_{\pm}$ are graphically displayed, it clearly appears that, while
$\bar{E}_{-}\cong E_{-}$, the other solution $\bar{E}_{+}$ is much larger than
$E_{+}$, the more so the greater $\epsilon$ is.
It is thus clear that the averaging distribution (\ref{eq:rho}), while 
mildly affecting the eigenvalue of $H$ lying in the $P$-space, drives away the
one lying in the $Q$-space.

\begin{figure}[p]
\begin{center}
\mbox{\epsfig{file=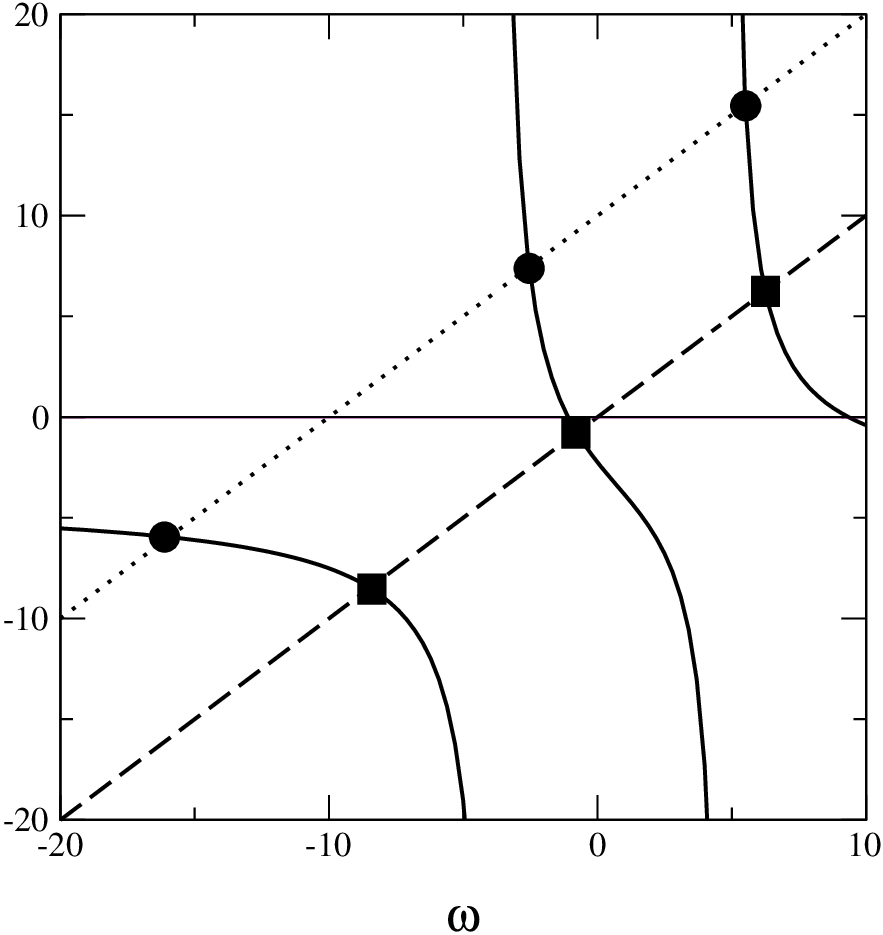,height=.28\textheight}}
\caption{ The eigenvalues of a tri-dimensional Hilbert space: The case of a
one-dimensional $P$-space. The following matrix elements of the Hamiltonian are
taken: $a_{11}=-4$, $a_{22}=-2$, $a_{33}=3$, $a_{12}=2$,
$a_{13}=-5$ and $a_{23}=3.5$, in arbitrary units.
The exact eigenvalues (squares) and the ones of the energy averaged Hamiltonian
$\bar{\cal H}$ (circles) are shown. They correspond to the intersections with
the straight line $E=\omega+\epsilon$: The former with $\epsilon=0$, the latter
with $\epsilon=10$. The stability of the $P$-space eigenvalue and the upward
shift of those belonging to the $Q$-space are clearly apparent. }
\label{fig:hyperbola2}
\mbox{\epsfig{file=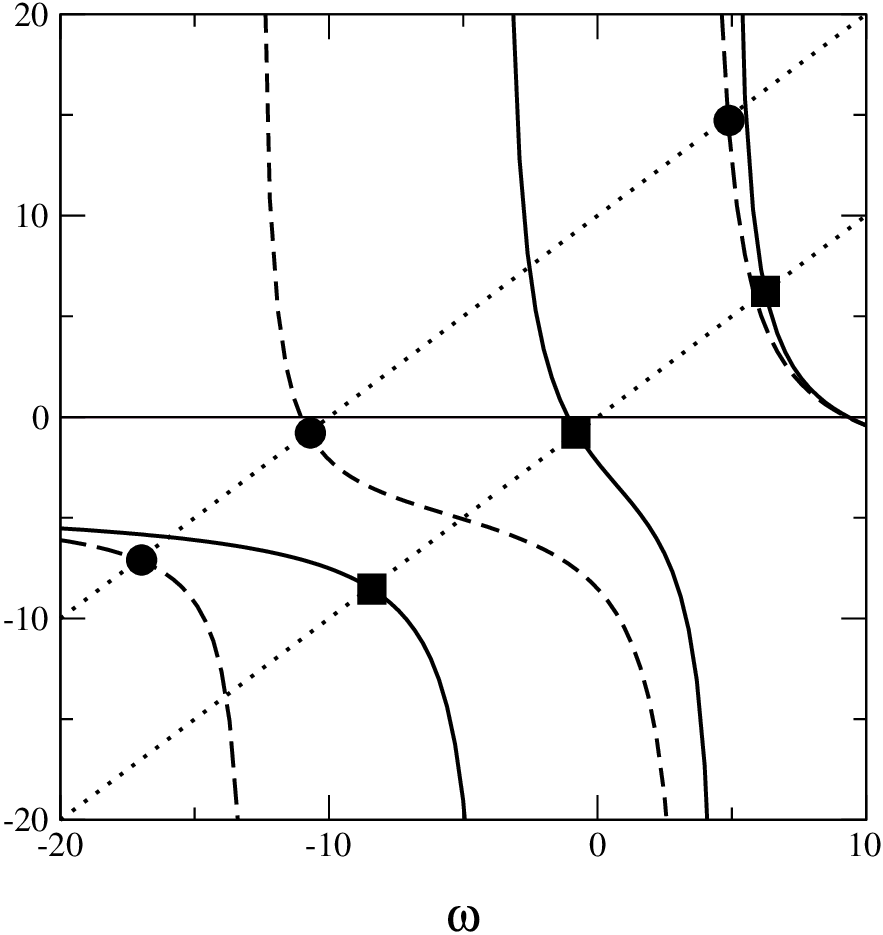,height=.28\textheight}}
\caption{ The eigenvalues of a tri-dimensional Hilbert space: The case of a 
bi-dimensional $P$-space. The matrix elements of the Hamiltonian are taken as 
in Fig.~\protect\ref{fig:hyperbola2}. The exact eigenvalues (squares) 
correspond to the intersections of the straight line $E=\omega$ with the
continuous curve.
The eigenvalues of the energy averaged Hamiltonian $\bar{\cal H}$ correspond to
the intersections of the straight line $\bar{E}=\omega+\epsilon$, with
$\epsilon=10$, with the dashed curves. Note the dependence upon $\epsilon$. 
The stability of the $P$-space eigenvalue and the upward shift of the one 
belonging to the $Q$-space are again clearly apparent. }
\label{fig:hyperbola3}
\end{center}
\end{figure}

\subsection{ Tri-dimensional Hilbert space }
\label{subsec:tri}

The space is spanned by the normalized states $|\chi_1\rangle$,
$|\chi_2\rangle$ and $|\chi_3\rangle$. Now, two choices are possible for the
projectors, namely
\begin{mathletters}
  \label{eq:part1}
\begin{eqnarray}
  P &\equiv& |\chi_1\rangle\langle\chi_1| + |\chi_2\rangle\langle\chi_2| \\
  Q &\equiv& |\chi_3\rangle\langle\chi_3|
\end{eqnarray}
\end{mathletters}
and
\begin{mathletters}
  \label{eq:part2}
\begin{eqnarray}
  P &\equiv& |\chi_1\rangle\langle\chi_1| \\
  \label{eq:part2b}
  Q &\equiv& |\chi_2\rangle\langle\chi_2| + |\chi_3\rangle\langle\chi_3|.
\end{eqnarray}
\end{mathletters}
In both cases, Eq.~(\ref{eq:Hcaleq}) can be recast as follows
\begin{equation}
  \label{eq:Ecub}
  (E-a_{11})(E-a_{22})(E-a_{33})-|a_{12}|^2(E-a_{33})-|a_{13}|^2(E-a_{22})
    -|a_{23}|^2(E-a_{11})=0,
\end{equation}
which is the cubic equation yielding the exact eigenvalues.
Note that Eq.~(\ref{eq:Ecub}) is easily obtained with the choice
(\ref{eq:part1}), because in this case the operator $(E-H_{QQ})^{-1}$ is
expanded in terms of the idempotent operator $|\chi_3\rangle\langle\chi_3|$.
Not so with the choice (\ref{eq:part2}), because now $(E-H_{QQ})^{-1}$ should
be expanded in terms of the operator (\ref{eq:part2b}), {\em which is not
idempotent}. Actually, the larger the powers of the latter are, the more
cumbersome they become. Yet, also in this case it can be proved that
Eq.~(\ref{eq:Ecub}) holds valid.

Let us now examine the solutions of Eq.~(\ref{eq:Hbarav}): As in the previous
bi-dimensional case it is convenient to display the solutions graphically.
For the partition (\ref{eq:part2}), one finds that they are given by
the intersections of the $\epsilon$-independent curve
\begin{equation}
  \label{eq:EbarB}
  \bar{E} = a_{11} + \frac{|a_{12}|^2(\omega-a_{33})}{{\cal D}_1(\omega)}
    + \frac{|a_{13}|^2(\omega-a_{22})}{{\cal D}(\omega)}
    + \frac{a_{12}a_{13}^{*}a_{23}}{{\cal D}(\omega)}
    + \frac{a_{12}^{*}a_{13}a_{23}^{*}}{{\cal D}(\omega)},
\end{equation}
where
\begin{equation}
  \label{eq:Dcal}
  {\cal D}(\omega) = (\omega-a_{22})(\omega-a_{33})-|a_{23}|^2,
\end{equation}
with the straight line $\bar{E}=\omega+\epsilon$, as displayed in
Fig.~\ref{fig:hyperbola2}, where the case $\epsilon=0$, --- which clearly
provides the exact eigenvalues $E_i$ of the Schroedinger equation, --- is also
shown. From the figure, it transparently appears that $\bar{E}_0\cong E_0$,
whereas $\bar{E}_1>>E_1$ and $\bar{E}_2>>E_2$, the latter inequalities being
stronger when the parameter $\epsilon$ is large.

In the case of the partition (\ref{eq:part1}), the solutions are given by
the intersections of the curve obtained replacing $a_{22}\to a_{22}-\epsilon$ 
in Eqs.~(\ref{eq:EbarB}) and (\ref{eq:Dcal}) with the straight line
$\bar{E}=\omega+\epsilon$, as displayed in Fig.~\ref{fig:hyperbola3}.
We face here a new situation, since now not only the straight line, but also
Eqs.~(\ref{eq:EbarB}) and (\ref{eq:Dcal}) are $\epsilon$-dependent.
Yet, one again sees that for $\epsilon=0$ one recovers the eigenvalues $E_i$,
whereas when $\epsilon\ne0$ the intercepts occur for $\bar{E}_0\cong E_0$ and
$\bar{E}_1\cong E_1$, but for $\bar{E}_2>>E_2$.
Hence, we conclude that the action of the averaging distribution (\ref{eq:rho})
affects very little the eigenvalues belonging to the $P$-space, while pushing
off the ones in the $Q$-space by an amount proportional to $\epsilon$.

\section{ Enlarging the $P$-space }
\label{sec:Pspace}

As in our previous work, we address in the following the problem of the ground
state of nuclear matter, again shortly summarizing below the equations derived 
in Refs.~\cite{Car98,DeP99}). Indeed, the improvements brought in by the 
present research are more transparently appreciated by performing the analysis 
borning in mind the one formerly developed.

As a preliminary in carrying out our task, it is necessary to define the
operators $P$ and $Q$. For this purpose the natural candidates as building
blocks of the $P$ operator appear to be the eigenstates $|\phi_n\rangle$ of the
mean field Hamiltonian (\ref{eq:Hbarav}), defined by the equation 
\begin{equation}
  \label{eq:Hbareq}
  \bar{\cal H} |\phi_n\rangle = \bar{E}_n |\phi_n\rangle.
\end{equation}
Their finding requires, however, the solution of a difficult self-consistency 
problem.
Hence, in Refs.~\cite{Car98,DeP99} the simpler choice had been made of viewing 
as building blocks of $P$ the Hartree-Fock (HF) variational solutions, which
are trivial in nuclear matter.

Furthermore, there, for sake of illustration, the severe limitation of a 
one-dimensional $P$-space was adopted by setting
\begin{equation}
  \label{eq:Pone}
  P = |\chi_{\text{HF}}\rangle\langle\chi_{\text{HF}}|,
\end{equation}
$|\chi_{\text{HF}}\rangle$ being the HF ground state wave function of nuclear
matter (the Fermi sphere).
It was then shown that, on the basis of (\ref{eq:Pone}), one derives the mean
field equation
\begin{equation}
  \label{eq:E0barone}
  \bar{E}_0 = E_{\text{HF}} + \frac{\beta^2}{\bar{E}_0-\epsilon-E},
\end{equation}
which relates the mean field ($\bar{E}_0$), the HF ($E_{\text{HF}}$) and the
true ($E$) energies {\em per particle}, and the fluctuation equation
\begin{equation}
  \label{eq:fluctone}
  E-\bar{E}_0 = \pm\frac{1}{E-\bar{\epsilon}_2}\sqrt{\frac{2}{{\cal N}_2}}
  \beta^2, 
\end{equation}
where
\begin{equation}
  \label{eq:beta2}
  \beta^2 = 
  \sum_{\text{2p-2h}}|\langle\psi_{\text{2p-2h}}|V|\chi_{\text{HF}}\rangle|^2,
\end{equation}
the bras $\langle\psi_{\text{2p-2h}}|$ representing the two-particle--two-holes
(2p-2h) states of nuclear matter, whose average energy {\em per particle} is
$\bar{\epsilon}_2$. 
We emphasize that all the quantities appearing in Eqs.~(\ref{eq:E0barone}) and
(\ref{eq:fluctone}) are {\em per particle}, including the parameter $\epsilon$
and the residual effective interaction $V$, both of which should accordingly be
thought of here as being divided by the nuclear mass number $A$.

Note also that Eq.~(\ref{eq:fluctone}) gives the ``error'' of the mean field
energy in the first order of the complexity expansion: Hence, it is contributed
to only by the sector of the $Q$-space set up with the 2p-2h excitations.
Actually, again for sake of simplicity, the states of the $Q$-space, although
obeying well-defined, coupled differential equations (see Ref.~\cite{Car98}), 
have been assumed to be adequately represented by the HF 
multi-particle--multi-hole solutions, an approximation not impairing the 
orthogonality constraint $P\cdot Q=0$.

The sum in Eq.~(\ref{eq:beta2}) is performed over the ensemble of the 2p-2h
excited states lying in an appropriate energy range (in
Refs.~\cite{Car98,DeP99} taken to be fixed essentially by the parameter
$\epsilon$), whose number ${\cal N}_2$ can be computed using the Ericson's
formula \cite{Eri60} for the density of the spin $J$ $N$-particle--$N$-hole
nuclear states, namely
\begin{equation}
  \label{eq:rhoN}
  \rho^{(N)}_{ph}({\cal E},J) = \frac{g(g{\cal E})^{N-1}}{p!h!(N-1)!}
    \frac{2J+1}{\sqrt{8\pi}\sigma^3 N^{3/2}} \exp[-(2J+1)^2/(8N\sigma^2)] ,
\end{equation}
where
\begin{equation}
  \label{eq:gsigma}
  g = \frac{3}{2}\frac{A}{\epsilon_F} 
    \qquad \text{and} \qquad
  \sigma^2 = {\cal F}\sqrt{\frac{\cal E}{a}}\frac{1}{\hbar^2},
\end{equation}
${\cal F}$ being the nuclear moment of inertia, $\epsilon_F$ the Fermi energy,
$a=A/8$ MeV$^{-1}$ and ${\cal E}$ the excitation energy of the system.

Now, Eqs.~(\ref{eq:E0barone}) and (\ref{eq:fluctone}), because of the double
sign appearing in the latter, set up two systems, each one including two
equations, in two unknowns. Two options are possible in selecting the latter:
One can choose either the ground state mean field and true energies per
particle, i.~e. $\bar{E}_0$ and $E$, --- assuming the matrix elements of the
residual effective interaction to be under control, --- or $\bar{E}_0$ and
$\beta^2$, when $E$ is experimentally known, --- which is indeed the case in
nuclear matter.

The latter is the course we followed (and continue to follow in the present 
work) requiring the coincidence of the two $\bar{E}_0$ obtained by solving the 
two systems separately.
Actually, and notably, both systems lead to the same formal expression for the
mean field energy per particle, namely
\begin{equation}
  \label{eq:E0barsol}
  \bar{E}_0 = \frac{1}{2} \left\{(E_{\text{HF}}+E+\epsilon) -
    \sqrt{(E_{\text{HF}}-E-\epsilon)^2+4\beta^2} \right\} ,
\end{equation}
which holds valid for $\epsilon<E_{\text{HF}}-E$, the right hand side of this
inequality being positive because of the variational principle.
The above, when $\beta^2\to0$, yields
\begin{equation}
  \bar{E}_0 = E + \epsilon,
\end{equation}
in accord with (\ref{eq:VPQP}), but also with (\ref{eq:fluctone}), which, for
$V\to0$, gives $\bar{E}_0=E$, entailing that when the residual effective
interaction vanishes no fluctuations occur and hence the parameter $\epsilon$
should vanish as well.

On the other hand, the two systems yield two different expressions for the sum 
of the matrix elements of the residual effective interaction squared, i.~e. 
\begin{mathletters}
\label{eq:belu}
\begin{eqnarray}
  \label{eq:b2l}
  \beta^2_l &=& \sqrt{{\cal N}_2/2} \frac{E_l-\bar{\epsilon}_2}{2} 
    \left\{ \left[E_l(1+\sqrt{{\cal N}_2/2})
    -(\epsilon+\sqrt{{\cal N}_2/2}\bar{\epsilon}_2)-E_{\text{HF}}\right]
    \right. \nonumber \\ 
  && \qquad \left.
    + \sqrt{ \left[E_l(1+\sqrt{{\cal N}_2/2})
    -(\epsilon+\sqrt{{\cal N}_2/2}\bar{\epsilon}_2)
    - E_{\text{HF}}\right]^2 + 4\epsilon(E_l-E_{\text{HF}})} \right\}
\end{eqnarray}
and
\begin{eqnarray}
  \label{eq:b2u}
  \beta^2_u &=& \sqrt{{\cal N}_2/2} \frac{E_u-\bar{\epsilon}_2}{2} 
    \left\{ \left[E_u(-1+\sqrt{{\cal N}_2/2})
    -(-\epsilon+\sqrt{{\cal N}_2/2}\bar{\epsilon}_2)+E_{\text{HF}} 
    \right] \right. \nonumber \\
  && \qquad \left.
    +\sqrt{\left[E_u(-1+\sqrt{{\cal N}_2/2})
    -(-\epsilon+\sqrt{{\cal N}_2/2}\bar{\epsilon}_2)
    + E_{\text{HF}}\right]^2 + 4\epsilon(E_u-E_{\text{HF}})} \right\} ,
\end{eqnarray}
\end{mathletters}
both of which vanish in the limit $\epsilon\to0$, in accord with the previous
discussion.

Formula (\ref{eq:b2l}), solution of the first system of equations (the one 
with the ``+'' sign on the right hand side of (\ref{eq:fluctone})), yields the 
value of the {\em energy dependent} quantity (\ref{eq:beta2}) (here denoted by 
$\beta^2_l$) on the lower border of the energy band expressing the fluctuations
of the ground state energy $E$ (and thus encompassing $\bar{E}_0$).
Formula (\ref{eq:b2u}), solution of the second system of equations (the one
with the minus sign  on the right hand side of (\ref{eq:fluctone})), provides
instead (\ref{eq:beta2}) (here denoted by $\beta^2_u$) on the upper border of
the band (remember that $E-\bar{\epsilon}_2<0$).

Of course, of the energy per particle $E$ we only know the experimental value,
not the values on the borders of the band: As a consequence, we can only 
surmise the width $W$ of the latter, thus
providing two different inputs for the energy $E$ appearing on the right hand
side of (\ref{eq:belu}), namely $E_u=E+W/2$ and $E_l=E-W/2$.
However, we can then explore whether,for a given $W$, a
value for the parameter $\epsilon$ can be found (not too large, not too small)
such to have the two mean field energies per particle to coincide.
If and when this search succeeds, then an orientation on $W$ (or, equivalently,
on the size of the fluctuations of the ground state energy) can be gained.

It is of importance to realize that the very existence of this scheme stems 
from the occurrence of the same quantity $\beta^2$ in both (\ref{eq:E0barone}) 
and (\ref{eq:fluctone}). This, in turn, is due to an approximation introduced 
in Ref.~\cite{DeP99}: It will be further discussed in Appendix~\ref{app:A}.

We now improve upon the above approach by first letting the projector $P$ to 
encompass, beyond the ground state, the 2p-2h excitations of the HF variational
scheme as well. Thus, instead of Eq.~(\ref{eq:Pone}), we shall write
\begin{equation}
  \label{eq:P2}
  P = |\chi_{\text{HF}}\rangle\langle\chi_{\text{HF}}| + 
  \sum_\beta|\chi_{\text{HF}}^{2\beta}\rangle\langle\chi_{\text{HF}}^{2\beta}|,
\end{equation}
where the sum is meant to be extended to the whole set of 2p-2h HF excitations
$|\chi_{\text{HF}}^{2\beta}\rangle$.

With the choice (\ref{eq:P2}) the mean field Hamiltonian (\ref{eq:Hbarav}) is
then defined and one can compute the mean field ground state energy per
particle, $\bar{E}_0 = \langle\phi_0|\bar{\cal H}|\phi_0\rangle$, using for the
ket $|\phi_0\rangle$ the expression
\begin{equation}
  \label{eq:phi02}
  |\phi_0\rangle = s_0|\chi_{\text{HF}}\rangle + 
    \sum_{\gamma}s_2^\gamma|\chi_{\text{HF}}^{2\gamma}\rangle,
\end{equation}
and accounting for the influence of the $Q$-space on the ground state mean
energy per particle in first order of the complexity expansion, i.~e. by
setting 
\begin{equation}
  \label{eq:Q4}
  Q = \sum_{\gamma} 
    |\chi_{\text{HF}}^{4\gamma}\rangle\langle\chi_{\text{HF}}^{4\gamma}|,
\end{equation}
where the sum runs over the whole set of the HF 4p-4h excitations.
In (\ref{eq:phi02}) $s_0$ and $s_2^\gamma$ are complex coefficients, fixed,
in principle, by Eq.~(\ref{eq:Hbareq}) and satisfying the normalization 
condition
\begin{equation}
  |s_0|^2+\sum_{\gamma}|s_2^\gamma|^2 = 1.
\end{equation}
After straightforward, but lengthy, algebra using (\ref{eq:phi02}) and
(\ref{eq:Q4}) one finally arrives at the following new mean field equation
\begin{eqnarray}
  \label{eq:E0bartwoA}
  \bar{E}_0 &=& |s_0|^2 E_{\text{HF}} + 
    \sum_{\gamma}|s_2^\gamma|^2
    \langle\chi_{\text{HF}}^{2\gamma}|H|\chi_{\text{HF}}^{2\gamma}\rangle +
    2s_0^*\sum_{\gamma}s_2^\gamma
    \langle\chi_{\text{HF}}|{\cal V}|\chi_{\text{HF}}^{2\gamma}\rangle +
    \nonumber \\
  && +\frac{1}{\bar{E}_0-\epsilon-E}\sum_{\beta\gamma}|s_2^\gamma|^2
    |\langle\chi_{\text{HF}}^{4\beta}|V|\chi_{\text{HF}}^{2\gamma}\rangle|^2,
\end{eqnarray}
${\cal V}$ being the bare NN potential.

Notably, the above equation turns out to formally coincide with
(\ref{eq:E0barone}). Indeed, one recognizes in the first
three terms on the right hand side of Eq.~(\ref{eq:E0bartwoA}) just the mean 
value of the original bare Hamiltonian $H$ in the state $|\phi_0\rangle$, given
in (\ref{eq:phi02}).
In the thermodynamic limit characterizing
nuclear matter what survives out of these pieces is just the HF energy, since
the correction to the latter due to a finite number of particle-hole 
excitations goes to zero when divided, --- in order to get the energy per 
particle, --- by the infinite number of particles in the system: In other 
words, the 2p-2h admixture into (\ref{eq:phi02}) {\em does not change} the 
expectation value of the Hamiltonian.

Hence, defining 
\begin{equation}
  \label{eq:zeta2}
  \zeta^2 = \sum_{\beta\gamma} |s_2^\gamma|^2
    |\langle\chi_{\text{HF}}^{4\beta}|V|\chi_{\text{HF}}^{2\gamma}\rangle|^2,
\end{equation}
(\ref{eq:E0bartwoA}) can be recast into the form
\begin{equation}
  \label{eq:E0bartwo}
  \bar{E}_0 = E_{\text{HF}} + \frac{\zeta^2}{\bar{E}_0-\epsilon-E},
\end{equation}
whose similarity with Eq.~(\ref{eq:E0barone}) is transparent.

The same will take place for {\em any} admixture of $N$p-$N$h HF excited states
in $|\chi_{\text{HF}}\rangle$: Hence, in our framework different choices of the
projection operator $P$ lead to the same structure for the mean field equation
for nuclear matter. This invariance does not hold in finite nuclei.

The only, of course important, difference between (\ref{eq:beta2}) and 
(\ref{eq:zeta2}) is that in the latter the residual interaction $V$ induces 
transitions from 2p-2h to 4p-4h states, rather than from the Fermi sphere to 
the 2p-2h states.

Concerning the fluctuation equation one can again use (\ref{eq:E-E0}),
with the state $|\phi_0\rangle$ given now by Eq.~(\ref{eq:phi02}).
Then invoking the randomness of the phases of the $Q$-space wave functions
(RPA) and proceeding exactly as previously done in \cite{Car98,DeP99}, one
deduces the new fluctuation equation 
\begin{equation}
  \label{eq:flucttwo}
  E-\bar{E}_0 = \pm\frac{1}{E-\bar{\epsilon}_4}\sqrt{\frac{2}{{\cal N}_4}}
  \zeta^2, 
\end{equation}
where $\bar{\epsilon}_4$ denotes the average energy per particle of the 4p-4h 
HF states.
In (\ref{eq:flucttwo}), ${\cal N}_4$ represents the number of 4p-4h excitations
contributing to the sum over the index $\beta$ in (\ref{eq:zeta2}).

Hence, the ``formal invariance'', with
respect to the choice for the projector $P$, holds for both the equations at
the core of our statistical approach to nuclear matter in first order of the
complexity expansion.

Indeed, the inclusion of $N$p-$N$h states (with $N>2$) into the $P$-space would
merely imply the replacement, in (\ref{eq:flucttwo}), of ${\cal N}_4$ with 
${\cal N}_{N+2}$ and, at the same time, to have $\zeta^2$ defined in terms of
the matrix elements of $V$ between $N$p-$N$h and $(N+2)$p-$(N+2)$h states.
In addition, one should of course insert in the energy denominator the average
energy per particle of the $(N+2)$p-$(N+2)$h HF states.

Therefore, in this connection one clearly sees that the extension of the 
$P$-space rapidly leads to the vanishing of the fluctuations, owing to the very
fast increase of the number ${\cal N}_N$.

However, it should be realized that Eq.~(\ref{eq:E-E0}), as it stands, {\em is 
not} in general a good starting point to derive the fluctuation equation.
Indeed, it selects out only {\em one term} in the spectral decomposition of the
operator $1/(E-\bar{\cal H})$, which, as already pointed out, is justified for 
a one-dimensional $P$-space, but not so for a multidimensional one. 
Hence, we depart from
our past work \cite{Car98,DeP99}, not only by employing Eq.~(\ref{eq:P2}), but 
also by writing, in place of (\ref{eq:spect}), 
\begin{equation}
  \label{eq:spect2}
  |P\psi\rangle = 
    |\phi_0\rangle\frac{\langle\phi_0|V_{PQ}|Q\psi\rangle}{E-\bar{E}_0} +
    |\phi_2\rangle\frac{\langle\phi_2|V_{PQ}|Q\psi\rangle}{E-\bar{E}_2} +
    \left(\frac{1}{E-\bar{\cal H}}\right)^{\prime\prime} V_{PQ}|Q\psi\rangle,
\end{equation}
with an obvious meaning of the double primed operator in the last term on the
right hand side.
Clearly, Eq.~(\ref{eq:spect2}) does not follows directly from (\ref{eq:P2}),
but it assumes that in the spectral decomposition of the
$(E-\bar{\cal H})^{-1}$ operator only one prominent (collective) state
$|\phi_2\rangle$ enters beyond the ground state $|\phi_0\rangle$.

Sticking to this model, instead of Eq.~(\ref{eq:Qpsipsi}), we likewise use
\begin{equation}
  \label{eq:Qpsipsi2}
  |Q\psi\rangle = \frac{1}{E-h^{(2)}_{QQ}}V_{QP}\left\{
    |\phi_0\rangle\langle\phi_0|P\psi\rangle +
    |\phi_2\rangle\langle\phi_2|P\psi\rangle\right\},
\end{equation}
being
\begin{equation}
  \label{eq:hQQ2}
  h^{(2)}_{QQ} = H_{QQ} + 
    V_{QP}\left(\frac{1}{E-\bar{\cal H}}\right)^{\prime\prime}V_{PQ}.
\end{equation}
Then, by left multiplying (\ref{eq:spect2}) (with $E=E_0$) by $\langle\phi_0|$ 
and using (\ref{eq:Qpsipsi2}), we obtain
\begin{equation}
  \label{eq:E-E02}
  E_0-\bar{E}_0 = 
    \langle\phi_0|V_{PQ}\frac{1}{E_0-h^{(2)}_{QQ}}V_{QP}|\phi_0\rangle +
    \langle\phi_0|V_{PQ}\frac{1}{E_0-h^{(2)}_{QQ}}V_{QP}|\phi_2\rangle
    \frac{\langle\phi_2|P\psi\rangle}{\langle\phi_0|P\psi\rangle},
\end{equation}
which generalizes Eq.~(\ref{eq:E-E0}).

In a perfectly analogous fashion, by left multiplying (\ref{eq:spect2}) (with 
$E=E_2$) by $\langle\phi_2|$ and using (\ref{eq:Qpsipsi2}), we obtain
\begin{equation}
  \label{eq:E-E2}
  E_2-\bar{E}_2 = 
    \langle\phi_2|V_{PQ}\frac{1}{E_2-h^{(2)}_{QQ}}V_{QP}|\phi_2\rangle +
    \langle\phi_2|V_{PQ}\frac{1}{E_2-h^{(2)}_{QQ}}V_{QP}|\phi_0\rangle
    \frac{\langle\phi_0|P\psi\rangle}{\langle\phi_2|P\psi\rangle}.
\end{equation}
In the above, $E_0$ and $E_2$ stand for the first two exact eigenvalues
of the Schroedinger equation; $\bar{E}_0$ and $\bar{E}_2$ for the corresponding
quantities associated with Eq.~(\ref{eq:Hbareq}). 
Of course, Eq.~(\ref{eq:E-E2}) has no counterpart in our previous work: Here, 
it shows that, in our frame, all the energies of the $P$-space fluctuate.

Now, the energy averaging of (\ref{eq:E-E02}) and (\ref{eq:E-E2}) vanishes by
definition, but the energy averaging of their square, which yields the
``error'', does not. Hence, proceeding along the lines of
Refs.~\cite{Car98,DeP99}, we subtract on the right hand side of both equations
their average values, square the expressions thus obtained and make use of RPA,
keeping of our expansion in the complexity of the $Q$-space states the first
term only.
Next, we exploit the structure of $|\phi_2\rangle$, which, like
$|\phi_0\rangle$, must be normalized, orthogonal to $|\phi_0\rangle$ and of the
form
\begin{equation}
  \label{eq:phi2HF}
  |\phi_2\rangle = \sum_{\beta}c_2^\beta|\chi_{\text{HF}}^{2\beta}\rangle +
    c_0|\chi_{\text{HF}}\rangle .
\end{equation}
We thus finally arrive at the equation
\begin{equation}
  \label{eq:flucttwo0}
  E_0-\bar{E}_0 = \pm\frac{1}{E_0-\bar{\epsilon}_4}\sqrt{\frac{2}{{\cal N}_4}}
  (\zeta^2+r\xi^2), 
\end{equation}
where the further definition 
\begin{equation}
  \label{eq:xi2}
  \xi^2 = \sum_{\beta\gamma} {s_2^\gamma}^* c_2^\gamma
    |\langle\chi_{\text{HF}}^{4\beta}|V|\chi_{\text{HF}}^{2\gamma}\rangle|^2
\end{equation}
has been introduced in addition to (\ref{eq:zeta2}); moreover, we have set 
\begin{equation}
  \label{eq:r}
  r \equiv \frac{\langle\phi_2|P\psi\rangle}{\langle\phi_0|P\psi\rangle}.
\end{equation}
Likewise, for the energy of the 2p-2h state of the $P$-space one obtains the
fluctuation equation
\begin{equation}
  \label{eq:flucttwo2}
  E_2-\bar{E}_2 = \pm\frac{1}{E_2-\bar{\epsilon}_4}\sqrt{\frac{2}{{\cal N}_4}}
  (\eta^2+\frac{\xi^2}{r}), 
\end{equation}
where, naturally,
\begin{equation}
  \label{eq:eta2}
  \eta^2 = \sum_{\beta\gamma} |c_2^\gamma|^2
    |\langle\chi_{\text{HF}}^{4\beta}|V|\chi_{\text{HF}}^{2\gamma}\rangle|^2.
\end{equation}
We thus see that the fluctuation equations (\ref{eq:flucttwo0}) and
(\ref{eq:flucttwo2}) are actually coupled through the term (\ref{eq:xi2}).

Concerning the mean field equations, it is clear that, in the enlarged scheme
brought about by the projector (\ref{eq:P2}) (actually, by our drastic
modelling of the solutions of the mean field Hamiltonian (\ref{eq:Hbarav}) with
$P$ given by (\ref{eq:P2})) and by the $|P\psi\rangle$ given
in (\ref{eq:spect2}), an equation should exist also for
the energy of the 2p-2h state. It can be derived by computing 
$\bar{E}_2=\langle\phi_2|\bar{\cal H}|\phi_2\rangle$ and, notably, it turns 
out to read 
\begin{equation}
  \label{eq:E2bar}
  \bar{E}_2 = E^{(2)}_{\text{HF}} + \frac{\eta^2}{\bar{E}_2-\epsilon-E_2},
\end{equation}
$E^{(2)}_{\text{HF}}$ representing the HF energy per particle of the system
in the 2p-2h excited state.
To identify this one is not necessary, since we split $E^{(2)}_{\text{HF}}$ 
into a part associated with the HF ground state and a part
associated with the 2p-2h {\em excitation energies}, both per particle.
Since the latter vanishes in the thermodynamic limit, --- as previously noted
in commenting Eq.~(\ref{eq:E0bartwoA}), then
$E^{(2)}_{\text{HF}}=E_{\text{HF}}$ , a relation we expect to be approximately
fulfilled also in a heavy nucleus.

We conclude from the above analysis that our approach leads to a set of mean 
field equations, one for each of the states lying in the $P$-space: These 
equations, unlike the fluctuation ones, are not coupled.

\section{ Normalization and fluctuation of the $P$-space ground state wave 
function 
}
\label{sec:S}

In our framework one could identify the ground state spectroscopic 
factor $S$ with the square root of the norm of $|P\psi\rangle$, the system's 
ground state wave function projection in $P$-space. To work out an explicit 
expression for $S$ we commence by exploiting the completeness of the normalized
eigenstates of $\bar{\cal H}$. Hence we write
\begin{equation}
  \label{eq:Sdef}
  S^2\equiv\langle P\psi|P\psi\rangle
    =\sum_{n=0}^M\langle P\psi|\phi_n\rangle\langle\phi_n|P\psi\rangle
    =1-\langle Q\psi|Q\psi\rangle.
\end{equation}
Now, if we confine ourselves to set $M=1$, as in Refs.~\cite{Car98,DeP99}, 
then the Eq.~(\ref{eq:Qpsipsi}) for $|Q\psi\rangle$ is
warranted and we rewrite (\ref{eq:Sdef}) as follows
\begin{equation}
  \label{eq:Sint1}
  S^2 = 1-\langle\phi_0|V_{PQ}\frac{1}{(E_0-h_{QQ})^2}V_{QP}|\phi_0\rangle
    |\langle\phi_0|P\psi\rangle|^2.
\end{equation}
Moreover, since the choice $M=1$ clearly entails 
\begin{equation}
  S^2 = |\langle\phi_0|P\psi\rangle|^2,
\end{equation}
by exploiting (\ref{eq:E-E0}) Eq.~(\ref{eq:Sint1}) can be recast into the form
\cite{Car98,DeP99}
\begin{equation}
  \label{eq:Sint2}
  S^2 = 1+S^2\left[\frac{d}{dE_0}(E_0-\bar{E}_0)
    +\frac{E_0-\bar{E}_0}{\bar{E}_0-\epsilon-E_0}\right].
\end{equation}
Finally, employing (\ref{eq:E0barsol}) the expression, previously obtained in
\cite{Car98,DeP99}, 
\begin{equation}
  \label{eq:Sfin}
  S^2 = \left[\frac{3}{2}+\frac{1}{2}
    \frac{E_{\text{HF}}-E_0-\epsilon-2d\beta^2/dE_0}
    {\sqrt{(E_{\text{HF}}-E_0-\epsilon)^2+4\beta^2}}
    +\frac{\epsilon}{\bar{E}_0-\epsilon-E_0}\right]^{-1}
\end{equation}
follows, in which the energy derivative of the sum of the square moduls of the 
vacuum--2p-2h matrix elements of the effective interaction appears (its 
explicit expression is given in Ref.~\cite{DeP99}). Note that (\ref{eq:Sfin}) 
goes to one as $\beta^2\to0$, as it should.

If, however, the expression (\ref{eq:Qpsipsi2}) for the $Q$-space wave function
is used, then, rather than (\ref{eq:Sint1}), one obtains for the ground state 
spectroscopic factor 
\begin{eqnarray}
  \label{eq:Sint21}
  S^2 &=& 1-|\langle\phi_0|P\psi\rangle|^2\langle\phi_0|
    V_{PQ}\frac{1}{(E_0-h_{QQ}^{(2)})^2}V_{QP}|\phi_0\rangle
    -|\langle\phi_2|P\psi\rangle|^2\langle\phi_2|
    V_{PQ}\frac{1}{(E_0-h_{QQ}^{(2)})^2}V_{QP}|\phi_2\rangle 
    \nonumber \\
  && -\langle\phi_0|P\psi\rangle \langle\phi_2|P\psi\rangle^*
    \langle\phi_2|
    V_{PQ}\frac{1}{(E_0-h_{QQ}^{(2)})^2}V_{QP}|\phi_0\rangle 
    \nonumber \\
  && -\langle\phi_0|P\psi\rangle^* \langle\phi_2|P\psi\rangle
    \langle\phi_0|
    V_{PQ}\frac{1}{(E_0-h_{QQ}^{(2)})^2}V_{QP}|\phi_2\rangle . 
\end{eqnarray}
The above, with the help of Eqs.~(\ref{eq:E-E02}) and (\ref{eq:E-E2}), and
assuming 
\begin{equation}
  \frac{1}{E_0-h_{QQ}^{(2)}} \approx \frac{1}{E_2-h_{QQ}^{(2)}},
\end{equation}
can be further elaborated and one ends up with the expression
\begin{equation}
  \label{eq:Sint22}
  S^2 = \frac{1-\left[
    \frac{\displaystyle d(\bar{E}_2-\bar{E}_0)}{\displaystyle dE_0}+
    \frac{\displaystyle\bar{E}_2-\bar{E}_0}
    {\displaystyle\bar{E}_0-\epsilon-E_0}\right]
    |\langle\phi_2|P\psi\rangle|^2}
    {1+\frac{\displaystyle\epsilon}{\displaystyle\bar{E}_0-\epsilon-E_0}
    +\frac{\displaystyle d\bar{E}_0}{\displaystyle dE_0}},
\end{equation}
which reduces to (\ref{eq:Sint2}), as it should, if
$\langle\phi_2|P\psi\rangle\to0$, i.~e. for a one-dimensional $P$-space.

Since, from Eq.~(\ref{eq:r}),
\begin{equation}
  \label{eq:phi2psi}
  |\langle\phi_2|P\psi\rangle|^2 = \frac{r^2}{1+r^2}S^2,
\end{equation}
then Eq.~(\ref{eq:Sint22}) can be recast as follows
\begin{equation}
  \label{eq:Sdef2}
  S^2 = \left\{1+\frac{1}{\bar{E}_0-\epsilon-E_0}\left[\epsilon+
    \frac{r^2}{1+r^2}\left(\bar{E}_2-\bar{E}_0\right)\right]+
    \frac{1}{1+r^2}\left(
    \frac{d\bar{E}_0}{dE_0}+r^2\frac{d\bar{E}_2}{dE_0}\right)\right\}^{-1},
\end{equation}
which again reduces to (\ref{eq:Sint2}) as $r\to0$.

We now address the problem of the fluctuation of $|P\psi\rangle$.
For this scope, we focus on the ground state and, by combining 
Eqs.~(\ref{eq:Hbar0eq}) and (\ref{eq:Schrav}), we obtain
\begin{equation}
  (E-\bar{\cal H})[|P\psi\rangle-|\biglb\langle P\psi\bigrb\rangle\rangle]
    +(E-\bar{E}_0)|\biglb\langle P\psi\bigrb\rangle\rangle =
    V_{PQ}|Q\psi\rangle
\end{equation}
(the angle brackets meaning energy averaging).

Then, if use is made of the expression (\ref{eq:Qpsipsi}) for $|Q\psi\rangle$ 
and of the spectral decomposition of the operator $(E-\bar{\cal H})^{-1}$, one 
gets 
\begin{eqnarray}
  \label{eq:fluct1}
  |P\psi\rangle-|\biglb\langle P\psi\bigrb\rangle\rangle &=& 
    {\sum}^\prime|\phi_n\rangle\frac{1}{E-\bar{E}_n}\langle\phi_n|V_{PQ}
    \frac{1}{E-h_{QQ}}V_{QP}|\phi_0\rangle\langle\phi_0|P\psi\rangle
    \nonumber \\
  && + \frac{1}{E-\bar{E}_0}|\phi_0\rangle\langle\phi_0|V_{PQ}
    \frac{1}{E-h_{QQ}}V_{QP}|\phi_0\rangle\langle\phi_0|P\psi\rangle
    \nonumber \\
  && - (E-\bar{E}_0)\sum_n\frac{1}{E-\bar{E}_n}|\phi_n\rangle\langle\phi_n|
    \biglb\langle P\psi\bigrb\rangle\rangle.
\end{eqnarray}
Now, since 
$|\biglb\langle P\psi\bigrb\rangle\rangle\propto|\phi_0\rangle$, from the above
finally it follows 
\begin{eqnarray}
  \label{eq:fluct2}
  |P\psi\rangle-|\biglb\langle P\psi\bigrb\rangle\rangle &=& 
    (1-|\phi_0\rangle\langle\phi_0|)^{-1}
    \left(\frac{1}{E-\bar{\cal H}}\right)^\prime V_{PQ}\frac{1}{E-h_{QQ}}
    V_{QP}|\phi_0\rangle\langle\phi_0|P\psi\rangle \nonumber \\
  &=& \langle\phi_0|P\psi\rangle
    \left(\frac{1}{E_0-\bar{\cal H}}\right)^\prime 
    V_{PQ}\frac{1}{E_0-h_{QQ}}V_{QP}
    |\phi_0\rangle,
\end{eqnarray}
which {\em vanishes when $P$ is given by (\ref{eq:Pone})}, since clearly 
$|P\psi\rangle$ does not fluctuate in a one-dimensional $P$-space.

If, on the other hand, $P$ is given by Eq.~(\ref{eq:P2}), then
the above can be computed, in first order of the complexity expansion, using
(\ref{eq:phi02}) for $|\phi_0\rangle$, (\ref{eq:phi2HF})
for $|\phi_2\rangle$ and (\ref{eq:Q4}) for $Q$. One ends up with the expression
(we set $E=E_0$ to conform to previous notations) 
\begin{equation}
  [|P\psi\rangle-|\biglb\langle P\psi\bigrb\rangle\rangle]_1 = 
    \langle\phi_0|P\psi\rangle\frac{|\phi_2\rangle}{E_0-\bar{E}_2}
    \sum_{\beta\beta'\gamma}
    \langle\chi_{\text{HF}}^{2\beta}|V|\chi_{\text{HF}}^{4\gamma}\rangle
    \frac{c^*_{2\beta}s^{\phantom{*}}_{2\beta'}}
    {E_0-\epsilon_{\text{HF}}^{4\gamma}}
    \langle\chi_{\text{HF}}^{4\gamma}|V|\chi_{\text{HF}}^{2\beta'}\rangle,
\end{equation}
which can be further simplified invoking the randomness of the phases of the
wave functions in the $Q$-space and again introducing the average energy per
particle $\bar{\epsilon}_4$ for the 4p-4h HF excited states. 
Then, with the help of Eq.~(\ref{eq:xi2}), the formula
\begin{equation}
  \label{eq:psiflucttwo}
  [|P\psi\rangle-|\biglb\langle P\psi\bigrb\rangle\rangle]_1 =
    \frac{\langle\phi_0|P\psi\rangle}
    {(E_0-\bar{E}_2)(E_0-\bar{\epsilon}_4)}|\phi_2\rangle\xi^2
\end{equation}
is derived. It gives the fluctuations of the wave function in first order of 
the complexity expansion.

Notice that here the scalar product $\langle\phi_0|P\psi\rangle$, unlike in 
Refs.~\cite{Car98,DeP99}, does not coincide with the spectroscopic factor
$S$, as defined in (\ref{eq:Sdef}), {since Eq.~(\ref{eq:psiflucttwo}) has been
obtained using an enlarged $P$-space. Rather, it measures the amount of
the true ground state wave function of the system embodied in the mean field 
state $|\phi_0\rangle$.

\section{ Results from a schematic model }
\label{sec:res}

The numerical implementation of the formalism described in the previous 
Sections represents a challenging task.

Therefore, as a first step toward this goal, we develop here a version of our
approach that, while simpler, has the merit of being amenable to numerical
results.

To start with, because to solve Eq.~(\ref{eq:Hbareq}), with the projector $P$ 
given by (\ref{eq:P2}), is a major task, we assume the two lowest solutions of
the latter to be of the form
\begin{mathletters}
  \label{eq:phi0phi2}
  \begin{equation}
  |\phi_0\rangle = s_0|\chi_{\text{HF}}\rangle 
    + s_2|\chi_{\text{HF}}^{\text{2p-2h}}\rangle
  \end{equation}
and
  \begin{equation}
  |\phi_2\rangle = -s_2|\chi_{\text{HF}}\rangle 
    + s_0|\chi_{\text{HF}}^{\text{2p-2h}}\rangle,
  \end{equation}
\end{mathletters}
where $s_0^2+s_2^2=1$. Of course, $s_2$ remains to be fixed.

This assumption corresponds to take in the sum on the right hand side of
(\ref{eq:P2}) one term only and places all the strength of the 2p-2h 
eigenstates of $\bar{\cal{H}}$ in a single mode. Then, by defining
\begin{equation}
  \label{eq:mu2}
  \mu^2 = \sum_\gamma|\langle\chi_{\text{HF}}^{4\gamma}|V
    |\chi_{\text{HF}}^{\text{2p-2h}}\rangle|^2,
\end{equation}
the expressions (\ref{eq:zeta2}), (\ref{eq:xi2}) and (\ref{eq:eta2}) can be
recast as follows
\begin{mathletters}
  \begin{eqnarray}
    \zeta^2 &=& s_2^2\mu^2 \\
    \xi^2   &=& s_2\sqrt{1-s_2^2}\mu^2 \\
    \eta^2  &=& (1-s_2^2)\mu^2,
  \end{eqnarray}
\end{mathletters}
where, beyond $\mu^2$, only the coefficient $s_2$ enters.

Accordingly, the mean field and the fluctuations equations now read
\begin{mathletters}
\label{eq:ebar02mod}
  \begin{eqnarray}
    \label{eq:ebar02moda}
    \bar{E}_0 &=& E_{\text{HF}} + \frac{s_2^2\mu^2}{\bar{E}_0-\epsilon-E_0} \\
    \label{eq:ebar02modb}
    \bar{E}_2 &=& E_{\text{HF}}^{(2)} +
      \frac{(1-s_2^2)\mu^2}{\bar{E}_2-\epsilon-E_2}
  \end{eqnarray}
\end{mathletters}
and
\begin{mathletters}
  \label{eq:flucteq2}
  \begin{eqnarray}
    \label{eq:flucteq2a}
    E_0-\bar{E}_0 &=& \pm\sqrt{\frac{2}{{\cal N}_4}}
      \frac{\mu^2}{E_0-\bar{\epsilon}_4}
      (s_2^2\pm r s_2\sqrt{1-s_2^2}) \\
    \label{eq:flucteq2b}
    E_2-\bar{E}_2 &=& \pm\sqrt{\frac{2}{{\cal N}_4}}
      \frac{\mu^2}{E_2-\bar{\epsilon}_4}
      (1-s_2^2\pm\frac{1}{r} s_2\sqrt{1-s_2^2}),
  \end{eqnarray}
\end{mathletters}
where $E_{\text{HF}}^{(2)}$ and $\bar{\epsilon}_4$ have been previously 
defined. We thus see that each dimension added to the $P$-space entails the
occurrence of {\em two} new systems in two equations.
Furthermore, the double sign inside the round brackets on the right hand side
of Eqs.~(\ref{eq:flucteq2}) corresponds to the two options one has in choosing
$s_0$, namely
\begin{equation}
  \label{eq:s0sign}
  s_0 = \pm\sqrt{1-s_2^2}.
\end{equation}

Although the above equations are strictly valid in the thermodynamic limit only
($A\to\infty$), we shall make an heuristic use of them also for the nucleus 
$A=208$, which is of course finite, but large.

Concerning the ground state spectroscopic factor (\ref{eq:Sdef2}), in the 
present model the energy derivative appearing in $S^2$ is easily found (from
(\ref{eq:ebar02mod})) to read
\begin{equation}
  \label{eq:der2}
    \frac{d\bar{E}_0}{dE_0} = \frac{1}{2} + \frac{1}{2} 
      \frac{E_{\text{HF}}-E_0-\epsilon-2d(s_2^2\mu^2)/dE_0}
      {\sqrt{(E_{\text{HF}}-E_0-\epsilon)^2+4s_2^2\mu^2}},
\end{equation}
where account is taken of the energy dependence of the coefficient $s_2$.

In accord with the general theory previously discussed, both $\bar{E}_0$ and
$\bar{E}_2$ lie inside an energy band expressing the ``error'' they are
affected by. In conformity, also the ground state spectroscopic factor should 
be computed on the lower and upper borders of the ground state energy band: 
Hence, we need to compute on both boundaries Eq.~(\ref{eq:der2}),
which in turn requires the knowledge of the energy derivative of the quantity
$(s_2\mu)^2$. 

For this purpose, in principle one should use the residual effective 
interaction on the lower boundary, which is found to be
\begin{mathletters}
  \label{eq:mu2lu}
  \begin{eqnarray}
    s_2^2\mu_l^2 &=& d_0 \frac{E_0^l-{\bar{\epsilon}_4}}{2}\Big\{
      E_0^l(1+d_0)-\epsilon-d_0{\bar{\epsilon}_4}-E_{\text{HF}}
      \nonumber \\ && \quad +\sqrt{
    [E_0^l(1+d_0)-\epsilon-d_0{\bar{\epsilon}_4}-E_{\text{HF}}]^2
      +4\epsilon(E_0^l-E_{\text{HF}})}\Big\},
  \end{eqnarray}
and on the upper one, where it reads
  \begin{eqnarray}
    s_2^2\mu_u^2 &=& d_0 \frac{E_0^u-{\bar{\epsilon}_4}}{2}\Big\{
    E_0^u(-1+d_0)+\epsilon-d_0{\bar{\epsilon}_4}+E_{\text{HF}}
      \nonumber \\ && \quad +\sqrt{
   [E_0^u(-1+d_0)+\epsilon-d_0{\bar{\epsilon}_4}+E_{\text{HF}}]^2
      +4\epsilon(E_0^u-E_{\text{HF}})}\Big\},
  \end{eqnarray}
\end{mathletters}
having set
\begin{equation}
  \label{eq:d0}
  d_0 = \sqrt{\frac{{\cal N}_4}{2}}\frac{1}{1+r\sqrt{1-s_2^2}/s_2}
\end{equation}
(the double sign stemming from (\ref{eq:s0sign})),
and where, as usual, $E_0^l$ and $E_0^u$ are the true ground state energies on
the lower and upper borders of the band: As already discussed, these are
expressed in terms of the band's width $W$.

However, in practice, owing to the energy dependence of {\em both} the residual
effective interaction $\mu$ and the coefficient $s_2$ appearing in the ground 
state mean field wave function, the energy derivative of (\ref{eq:mu2lu}) turns
out to be very cumbersome to carry out. Accordingly, we evaluated numerically
all the derivatives with respect to the exact ground state energy $E_0$
entering into the expression (\ref{eq:Sdef2}) for $S^2$.

Note that $d_0\to\sqrt{{\cal N}_4/2}$ when $r\to0$: One thus formally recovers 
in this limit the results of a one-dimensional $P$-space \cite{DeP99}.

In solving the systems of equations (\ref{eq:ebar02moda})--(\ref{eq:flucteq2a})
and (\ref{eq:ebar02modb})--(\ref{eq:flucteq2b}) we have chosen in
(\ref{eq:s0sign}) the plus sign. The other option need not be considered
because it will provide the same results, but for the change of sign of $s_2$.
Indeed, our equations are clearly invariant under the simultaneous
transformations $s_0\to-s_0$ and $s_2\to-s_2$, as it should be as the latter
simply correspond to a change of phase in the states (\ref{eq:phi0phi2}).

Concerning the other quantities needed in our equations, we have assumed for 
the binding energy per particle of nuclear matter 
\begin{mathletters}
\begin{equation}
  \label{eq:Esat}
  E_0=-16\,\text{MeV}
\end{equation}
for the ground state  and
\begin{equation}
  E_2=-14.8\, \text{MeV}
\end{equation}
\end{mathletters}
for the 2p-2h ``collective'' excited state.
The latter value is hinted by the calculations of the ($\gamma,nn$) and
($\gamma,pp$) photo-absorption cross-sections in nuclei \cite{Bof96} ($n$ being
a neutron, $p$ a proton).

For the HF energies per particle we have taken
\begin{mathletters}
  \begin{eqnarray}
    E_{\text{HF}} &=& -12\, \text{MeV} \\
    E^{(2)}_{\text{HF}} &=& -11.96\, \text{MeV} \\
    \bar{\epsilon}_4 &=& -11.92\, \text{MeV}.
  \end{eqnarray}
\end{mathletters}
The first of the above actually corresponds to a typical BHF outcome 
(see, e.~g., Ref.~\cite{Coe70}). The other two stem from the minimal energy
($\cong 4$ MeV) of a ph excitation in a Wood-Saxon well with parameters 
adjusted for the nucleus $Pb^{208}$ \cite{Pol01}.

The choice of a minimal ph energy has been made:
\begin{itemize}
\item[a)] because the only solution of our schematic model obtains with
  the residual effective interaction weakened by an order of magnitude with 
  respect to the bare one: Hence the latter can only effectively connect
  with the lowest 4p-4h excitations (see the results below);
\item[b)] because only with the minimal excitation energy of the 4p-4h states 
Eq.~(\ref{eq:rhoN}) provides a number of states in an energy range set by
$\epsilon$ still compatible with non-trivial fluctuation equations
(\ref{eq:flucteq2}). 
\end{itemize}

In connection with the point b) we have limited ourselves to use formula
(\ref{eq:rhoN}) for 4p-4h excited states with angular momentum $J=0$.
It should also be added that one can legitimately question the
validity of (\ref{eq:rhoN}), which is deduced in the framework of a Fermi gas
by purely combinatorial methods. 

With the energies per particle above quoted, --- which should clearly be 
viewed as merely orientative but for (\ref{eq:Esat}), --- we have solved the 
pair of systems corresponding to Eqs.~(\ref{eq:ebar02moda}) and 
(\ref{eq:flucteq2a}) and, as well, the one corresponding to 
Eqs.~(\ref{eq:ebar02modb}) and (\ref{eq:flucteq2b}), fulfilling the equal mean 
field constraint in both cases.
In addition, for sake of simplicity, we have required the nucleons' residual 
effective interaction $\mu$ to come out the same in both instances in accord 
with the schematic nature of our model.

Our findings for $\bar{E}_0$, $\bar{E}_2$, $\mu^2_l$ and $\mu^2_u$ and for the
ground state spectroscopic factor are displayed in Table~\ref{tab:I}.
These results have been obtained for a specific choice of $\epsilon$, which
fulfills the inequalities $0\leq\epsilon\leq E_{\text{HF}}-E=4$~MeV, $s_2$ and
$r$. Notably, in order to find a solution one has to restrict these parameters 
to a rather narrow range around the values quoted in the Table. Also the band
width $W$ cannot be much larger then 0.1 MeV in order to get a solution.
The numbers in Table~\ref{tab:I} correspond to the choice $W=0.1$ MeV.

\begin{table}[t]
  \begin{tabular}[p]{|cc|cc|cc|ccc|}
    $\bar{E}_0$ (MeV) & $\bar{E}_2$ (MeV) & $\mu^2_l$ (MeV$^2$) & 
    $\mu^2_u$ (MeV$^2$) & $S_l$ & $S_u$ & $s_2$ & $\epsilon$ (MeV) & $r$ \\
\tableline
    -16.001 & -15.901 & 25 & 26 & 1.016 & 1.000 & 0.71 & 3.2 & 1.02 \\
  \end{tabular}
  \caption{ The solution of the schematic model }
  \label{tab:I}
\end{table}

Although we consider the outcomes of this model as merely orientative and 
without any pretense of being in touch with the real physics, both because of 
the crudeness of the model itself and because of the uncertainties affecting 
our inputs, yet they are worth a few remarks.

First, the eigenstates of the mean field Hamiltonian $\bar{\cal H}$ appear to
have equal projections on the two vectors spanning the bi-dimensional $P$-space
of our model, as it follows from the values we have found for $s_2$.

Second, it is clear that, with respect to our past work with a one-dimensional
$P$-space, a reduction of the error affecting the mean field energy was
to be expected. It turned out to amount to an order of magnitude.
This fact points to a fast convergence of the complexity expansion or,
alternatively, to a contribution of the 6p-6h excitations to the binding energy
per particle (in nuclear matter) of less then 1\%.
This estimate is of course very rough: In fact, the magnitude of the error
arises not anly from the dimensions of the $P$-space, but as well from the
proximity of the HF (or BHF) solution to the experimental value of the binding
energy per particle.

Third, the massive reduction of the residual effective interaction as
compared to the bare one, already found in Refs.~\cite{Car98,DeP99}, is here
confirmed. Unfortunately, in our past work we had overlooked a factor of $A$
(the interaction under consideration is {\em per particle}): The quenching
there predicted was accordingly wrong, but still should be quantified in one 
order of magnitude (see Appendix~\ref{app:B}). 

Finally, a comment is in order on the spectroscopic factor $S$. For this
purpose we write, in terms of the states $|\chi_{\text{HF}}\rangle$ and 
$|\chi_{\text{HF}}^{\text{2p-2h}}\rangle$ (or, alternatively, $|\phi_0\rangle$
and $|\phi_2\rangle$) spanning our $P$-space,
\begin{eqnarray}
  \label{eq:Ppsimod}
  |P\psi\rangle &=& |\chi_{\text{HF}}\rangle\langle\chi_{\text{HF}}|\psi\rangle
    + |\chi_{\text{HF}}^{\text{2p-2h}}\rangle
    \langle\chi_{\text{HF}}^{\text{2p-2h}}|\psi\rangle \nonumber \\
  &=& |\phi_0\rangle\langle\phi_0|P\psi\rangle +
    |\phi_2\rangle\langle\phi_2|P\psi\rangle.
\end{eqnarray}
From the above and (\ref{eq:phi0phi2}) it follows easily
\begin{eqnarray}
  \label{eq:phi0phi2Ppsi}
  \langle\phi_0|P\psi\rangle &=& \langle\chi_{\text{HF}}|\psi\rangle s_0
    + \langle\chi_{\text{HF}}^{\text{2p-2h}}|\psi\rangle s_2 \nonumber \\
  && \\
  \langle\phi_2|P\psi\rangle &=& - \langle\chi_{\text{HF}}|\psi\rangle s_2
    + \langle\chi_{\text{HF}}^{\text{2p-2h}}|\psi\rangle s_0. \nonumber \\
\end{eqnarray}
Inverting these equations and exploiting (\ref{eq:r}), we obtain
\begin{mathletters}
\begin{eqnarray}
  \langle\chi_{\text{HF}}|\psi\rangle &=& \langle\phi_0|P\psi\rangle s_0
    - \langle\phi_2|P\psi\rangle s_2 = \langle\phi_0|P\psi\rangle (s_0-rs_2) \\
  \langle\chi_{\text{HF}}^{\text{2p-2h}}|\psi\rangle &=& 
    \langle\phi_0|P\psi\rangle s_2 + \langle\phi_2|P\psi\rangle s_0 =
    \langle\phi_0|P\psi\rangle (s_2+rs_0).
\end{eqnarray}
\end{mathletters}
Hence, with the values for $s_2$ and $r$ quoted in Table~\ref{tab:I}, we 
conclude that
\begin{equation}
  \label{eq:HF0}
  \langle\chi_{\text{HF}}|\psi\rangle \cong 0, \quad
    \langle\chi_{\text{HF}}^{\text{2p-2h}}|\psi\rangle \cong 1,
\end{equation}
an outcome we are hardly ready to view as realistic.

On the other hand, taking the minus sign inside the round brackets in the
fluctuation equations (\ref{eq:ebar02mod}) and also in (\ref{eq:d0}) (which
amounts to the substitution $s_2\to-s_2$), we get
\begin{equation}
  \label{eq:HF1}
  \langle\chi_{\text{HF}}|\psi\rangle \cong 1, \quad
    \langle\chi_{\text{HF}}^{\text{2p-2h}}|\psi\rangle \cong 0,
\end{equation}
which, while drastic , appears to provide a physical picture more consistent 
with the simple model we are employing.
In fact, the above results hold because $\langle\phi_0|P\psi\rangle\cong1$,
which follows from the normalization condition
\begin{equation}
  |\langle\phi_0|P\psi\rangle|^2+|\langle\phi_2|P\psi\rangle|^2=1
\end{equation}
and from the finding that $r\cong1$.
Clearly, both (\ref{eq:HF0}) and (\ref{eq:HF1}) lead to a value of one for the
spectroscopic factor $S$, that is the result we have found.

The two radically different solutions (\ref{eq:HF0}) and (\ref{eq:HF1}) can be
reconciled through Eq.~(\ref{eq:psiflucttwo}). The latter, indeed, when
implemented in our schematic model, yields for the fluctuations of the ground
state wave function
\begin{equation}
   [|P\psi\rangle-|\phi_0\rangle]_1 \cong |\phi_2\rangle.
\end{equation}
Likewise, by either simmetry arguments or explicit calculation, one can get for
the collective 2p-2h excited state
\begin{equation}
   [|P\psi_{\text{2p-2h}}\rangle-|\phi_2\rangle]_1 \cong |\phi_0\rangle.
\end{equation}
Thus, it appears that, in the schematic model presented in this Section, while
the fluctuation of the energy is puny, the fluctuation of the wave function is 
the largest possible. This correlation is worth being explored in more
realistic models.

In conclusion, it appears warranted to view the scalar product
$\langle\phi_0|P\psi\rangle$ --- which is informative on how
much of the true ground state wave function of the system is embodied in the
corresponding mean field one, --- as a better representative of $S$: In this 
last instance, we would obtain $S\cong0.7$.

\section{ Conclusions }
\label{sec:concl}

In this paper we have first shortly revisited the statistical theory of the
ground and excited collective states of general many-body systems developed in
Refs.~\cite{Fes96,Car98,DeP99}. 

Next we have presented an improved version of
the approach by allowing for a $P$-space with larger dimensions than in
Refs.~\cite{Car98,DeP99}. In applying this new framework to nuclear matter we
have obtained both the mean field energies and the fluctuations (the ``error'')
of both energies and wavefunctions. This has been done in first order
of the expansion in the complexity of the states of the $Q$-space, which, in
our scheme, hosts the random aspects of the nuclear dynamics. Notably, the
extension of the $P$-space allows to acknowledge the fast rate of convergence 
of the complexity expansion, due to the rapid growth of the nuclear level
density with the excitation energy.

Finally, as a first step toward the testing of the theory on the physics of the
atomic nuclei we have worked out a schematic model that renders the approach
amenable to numerical predictions.
While rough, the model has nevertheless helped us in clarifying basic
aspects of our statistical approach.

In conclusion, we feel the duty to state that the statistical approach here
presented germinated out of the ideas discussed by Herman Feshbach in
Ref.~\cite{Fes96}. It has been subsequently pursued by him in collaboration
with us in Refs.~\cite{Car98,DeP99} and in this work.

\acknowledgements
This work has been partially supported by the INFN-MIT ``Bruno Rossi'' Exchange
Program. Discussions with Prof.~H. A. Weidenm\"uller are gratefully
acknowledged. 

\appendix

\section{}
\label{app:A}

The derivation of the coupled equations at the basis of the present approach is
possible because the same quantity $\beta^2$ occurs in both (\ref{eq:E0barone})
and (\ref{eq:fluctone}). This occurrence stems from an approximation 
introduced in Ref.~\cite{DeP99}, namely 
\begin{eqnarray}
  \label{eq:arithapprox}
  &&\sum_{\beta\gamma}
    \Biglb\langle 
    \langle\psi_{2\beta}|V|\phi_0\rangle^*
    \langle\psi_{2\gamma}|V|\phi_0\rangle
    \Bigrb\rangle 
    \Biglb\langle 
    \langle\psi_{2\beta}|V|\phi_0\rangle
    \langle\psi_{2\gamma}|V|\phi_0\rangle^*
    \Bigrb\rangle \nonumber \\
  \quad &\approx& 2{\cal A}\left\{ \sum_{\beta\gamma}
    \Biglb\langle 
    \langle\psi_{2\beta}|V|\phi_0\rangle^*
    \langle\psi_{2\gamma}|V|\phi_0\rangle
    \Bigrb\rangle \right\} \sum_{\beta\gamma}
    \Biglb\langle 
    \langle\psi_{2\beta}|V|\phi_0\rangle^*
    \langle\psi_{2\gamma}|V|\phi_0\rangle
    \Bigrb\rangle .
\end{eqnarray}
In the above, the operator ${\cal A}$ corresponds to the arithmetic mean of the
quantity within the angle brackets over the set of the 2p-2h states in the
energy interval $\epsilon$ centered around $\bar{\epsilon}_2$ (actually, one
should take the real part of the right hand side of the previous equation:
However, for the sake of the argument we shall consider real quantities).

This approximation gives reasonable results if the matrix elements are roughly
of the same order of magnitude: This is the reason why the sum in
(\ref{eq:arithapprox}) is limited to an energy interval around the mean 2p-2h
energy where the matrix elements are relevant.

We give here a few calculable examples that prove the above statement 
\begin{eqnarray}
  \frac{N\sum_{n=1}^N \sin(n)^4}{2[\sum_{n=1}^N \sin(n)^2]^2} &=& 
    \left\{\begin{array}{ll} 0.725, & N=10 \\
                             0.746, & N=100 \\
                             0.750, & N=\infty
           \end{array} \right. \\
  \frac{N\sum_{n=1}^N n^2}{2[\sum_{n=1}^N n]^2} &=& 
    \left\{\begin{array}{ll} 0.636, & N=10 \\
                             0.663, & N=100 \\
                             2/3, & N=\infty
           \end{array} \right. \\
  \frac{N\sum_{n=1}^N \exp(-n)^2}{2[\sum_{n=1}^N \exp(-n)]^2} &=& 
    \left\{\begin{array}{ll} 2.311, & N=10 \\
                             23.11, & N=100 \\
                             \infty, & N=\infty
           \end{array} \right. .
\end{eqnarray}
Were the approximation exact, then the above ratios should evaluate to one.

\section{}
\label{app:B}

We report in the following the expression for $\beta^2$, --- the sum of the 
square of the vacuum $\to$ 2p-2h matrix elements, --- for the schematic 
interaction employed in Refs.~\cite{Car98,DeP99}, that is
\begin{equation}
 {\cal V}(r) = g_A \, \frac{e^{-\mu_A r}}{r}\, - 
   g_B\, \frac{e^{-\mu_B r}}{r} \, \frac{ 1 + P_x }{2} ,
\end{equation}
$P_x$ being the exchange operator in coordinate space.

Following the Appendix~B of Ref.~\cite{Car98}, $\beta^2$ can be recast as the
combination of direct and exchange contributions such as
\begin{eqnarray}
  \beta^2_{\text{dir}} &=& A \frac{\varrho}{4 k_F} \Biggl\{ 
    \frac{1}{\tilde \mu_A^2 - \tilde \mu_B^2} \left(\tilde \mu_A \arctan 
    \tilde \mu_A -\tilde \mu_B \arctan \tilde \mu_B \right ) \\
  && + \frac{9}{5} + \frac{17}{12} (\tilde \mu_A^2 + \tilde \mu_B^2 )
    + \frac{1}{4} (\tilde \mu_A^4 + \tilde \mu_B^4 ) +
    \frac{1}{4} \tilde \mu_A^2  \tilde \mu_B^2 \nonumber \\
  && +\frac{1}{4} \frac{1}{\tilde \mu_A^2 - \tilde \mu_B^2}  \left [
    \tilde \mu_B^3 (3 + \tilde \mu_B^2 )^2 \arctan 
    \left ( \frac{1}{\tilde \mu_B} \right ) 
    - \tilde \mu_A^3 (3 + \tilde \mu_A^2 )^2 \arctan
    \left ( \frac{1}{\tilde \mu_A} \right ) \right ] \Biggr\}
\end{eqnarray}
and
\begin{eqnarray}
  \beta^2_{\text{exc}} &=& A\frac{2}{\pi^4 \varrho} \int_0^\infty dq\, 
    \frac{q}{\mu_A^2 +q^2} \int_0^\infty dr \frac{e^{-\mu_B r}}{r^6} \sin(qr)
    \Bigl\{ [\sin(k_F r)-k_F r\cos(k_F r)]^2 \nonumber \\
  && + [\sin(k_F r) - \sin(r\sqrt{k_F^2-q^2}) - k_F r\cos(k_F r) +
    r\sqrt{k_F^2-q^2}\cos(r\sqrt{k_F^2-q^2})]^2 \Bigr\} ,
\end{eqnarray}
where $\tilde \mu_{A,B} = {\mu_{A,B}}/{2 k_F}$ and 
$\varrho={2k_F^3}/{3\pi^2}$.

In the latter, a minor error in the expression reported in
Ref.~\cite{Car98}, essentially of no numerical consequence, has been corrected.
More importantly, the necessity of normalizing $\beta^2$ to the mass number
$A$ had been overlooked in Refs.~\cite{Car98,DeP99}: Hence, the previously
found quenching of the residual effective interaction with respect to the bare 
one should also be reduced by a factor $A$. For example, taking $^{208}$Pb as
an orientation, the quenching would turn out to be of an order of magnitude.

\end{document}